# Harmonic Electron Cyclotron Maser Emission along the Coronal Loop

Mehdi Yousefzadeh[1], Yao Chen[1], Hao Ning[1], and Mahboub Hosseinpour[2]
[1] Institute of Frontier and Interdisciplinary Science and Institute of Space Sciences, Shandong University, Shandong, People's Republic of China
yaochen@sdu.edu.cn, m.yousefzadeh6@gmail.com
[2] Faculty of Physics, University of Tabriz, Tabriz, Iran

## Abstract

Efficient radiation at second and/or higher harmonics of $\Omega_{ce}$ has been suggested to circumvent the escaping difficulty of the electron cyclotron maser emission mechanism when it is applied to solar radio bursts, such as spikes. In our earlier study, we developed a three-step numerical scheme to connect the dynamics of energetic electrons within a large-scale coronal loop structure with the microscale kinetic instability energized by the obtained nonthermal velocity distribution and found that direct and efficient harmonic X-mode (X2 for short) emission can be achieved due to the strip-like features of the distribution. That study only considered the radiation from the loop top at a specific time. Here we present the emission properties along the loop at different locations and timings. We found that, in accordance with our earlier results, few to several strip-like features can appear in all cases, and the first two strips play the major role in exciting X2 and Z (i.e., the slow extraordinary mode) that propagate quasi-perpendicularly. For the four sections along the loop, significant excitation of X2 is observed from the upper two sections, and the strongest emission is from the top section. In addition, significant excitation of Z is observed for all loop sections, while there is no significant emission of the fundamental X mode. The study provides new insight into coherent maser emission along the coronal loop structure during solar flares.

*Unified Astronomy Thesaurus concepts:* Solar activity (1475); Solar corona (1483); Radio bursts (1339); Solar coronal radio emission (1993)

*Supporting material:* animation

## 1. Introduction

Solar radio bursts with very high brightness temperatures are suggested to be released via a coherent radiation mechanism. One such mechanism with direct amplification of electromagnetic waves (X and/or O modes), the electron cyclotron maser emission (ECME), is excited by energetic electrons with a positive gradient of their velocity distribution function (VDF) in strongly magnetized plasmas with $\omega_{pe}/\Omega_{ce} < 1$, where $\omega_{pe}$ is the plasma oscillation frequency, and $\Omega_{ce}$ is the electron gyrofrequency. The framework of ECME was first proposed by Twiss (1958), Gaponov (1959), and Schneider (1959). Wu & Lee (1979) made a breakthrough by considering the relativistic effect in the resonance condition.

When it is applied to solar radio bursts, such as spikes, this mechanism encounters the so-called escaping difficulty (Melrose & Dulk 1982; Sharma & Vlahos 1984; Melrose 1991; Melrose & Wheatland 2016). The difficulty arises if the fundamental X-mode (X1) emission with frequency around $\Omega_{ce}$ is the major radiating mode, since during its outward escape, it may get significantly absorbed by thermal plasmas at the second harmonic layer (where the mode frequency is equal to or close to twice the local gyrofrequency). Efficient harmonic emission (X2 or O2) via ECME has been considered as one possibility to circumvent the difficulty, since the corresponding absorption effect is drastically weakened (see Ning et al. 2021a, 2021b; Yousefzadeh et al. 2021).

To investigate whether harmonic emissions can be excited, Yousefzadeh et al. (2021) developed a three-step numerical scheme consisting of (1) the nonlinear force-free field extrapolation method (Wheatland et al. 2000; Wiegelmann 2004; Wiegelmann et al. 2006) deriving the large-scale magnetic topology of a solar active region (AR), which presents a loop structure for further study; (2) the guiding center (GC; Northrop 1963) method to simulate the transport process of millions of electrons along the loop, where the VDF can be revealed by monitoring a specific loop section; and (3) the particle-in-cell (PIC) method to simulate the kinetic instability driven by energetic electrons with the obtained VDFs. Such a scheme is a starting point to develop numerical techniques for the multiscale process of solar radio bursts, involving the electron transport within the large-scale coronal loop structure and the kinetic scale of microscopic kinetic instabilities.

In Yousefzadeh et al. (2021), only VDFs around the loop top at one specific time were considered. Interesting strip-like features of the electron VDFs are observed, which represent the main driver of further kinetic instability. They reported efficient excitation of the X2 mode via ECME, important to resolve its escaping difficulty. Various levels of the scattering effect (weak, moderate, and strong), characterized by different scattering timescales (5, 1, and 0.5 s), have been considered. It was found that the rate of energy conversion from energetic electrons to X2 can reach up to $2.9 \times 10^{-3}$ for the weak scattering case.

Note that Yousefzadeh et al. (2021) only investigated the ECME process from the loop-top region at a specific time (1 s postinjection). The emission property may vary along the loop structure and change with time. To get a complete view, it is necessary to investigate such temporal and spatial variations. This is the major aim of the present study.

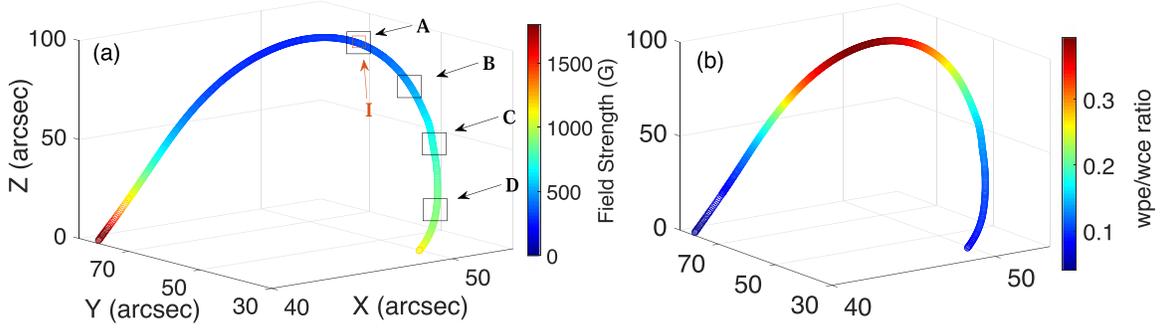

**Figure 1.** Selected magnetic field line derived from the HMI image (AR 11283). The left and right color bars represent the field strength and the $\omega_{pe}/\Omega_{ce}$ ratio, respectively. The letters A–D represent the sections within which VDFs will be examined, and the letter I represents the region of injection.

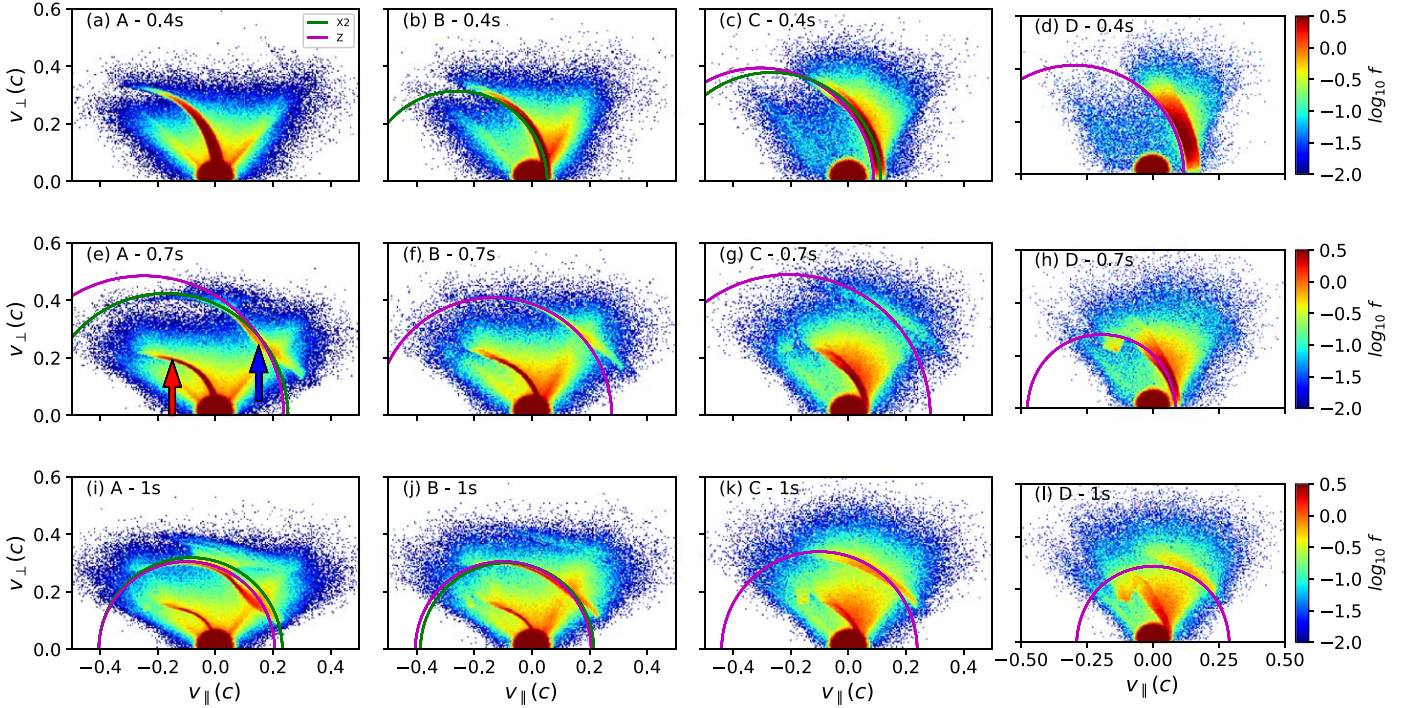

**Figure 2.** The VDFs at $t = 0\ \omega_{pe}^{-1}$ obtained by the PIC simulation for different sections (A, B, C, and D; left to right) along the loop structure at three different timings (0.4, 0.7, and 1 s postinjection; top to bottom). The red and blue arrows in panel (e) represent the first and second strips, respectively. Resonance curves (see Table 1 for corresponding parameters) and the background Maxwellian distribution are overplotted in each panel (the dark red region centered around the origin of the map). The video represents the temporal evolution of the VDFs within the four sections (A–D); it begins at $t = 0$ s and advances 0.1 s at a time until $t = 3$ s and then advances 2 s at a time until $t \sim 19$ s. The real-time duration of the video is 7 s (an animation relevant to this figure is available).
(An animation of this figure is available.)

## 2. Numerical Methods: The Three-step Scheme

The technical details of the three-step numerical scheme have been presented in Yousefzadeh et al. (2021). In the following, it is introduced briefly. To initiate the extrapolation, we select NOAA AR 11283, whose magnetogram data were observed by the Helioseismic and Magnetic Imager (HMI; Schou et al. 2012) at 21:30 UT on 2011 September 6. The same untwisted loop was selected for the present study. From the extrapolation, we can deduce the distribution of the magnetic field strength along the loop (see Figure 1(a)), and the plasma density along the loop is inferred with the following hydrostatic model (see, e.g., Newkirk 1961; Priest 1978):

$$n_{0e}(h) = n_0 \exp\left(-\frac{h}{H}\right), \quad (1)$$

where $H = 141$ Mm is the height scale (corresponding to $\sim 3$ MK), and the density $n_0$ is taken to be $10^9$ cm$^3$, appropriate for typical ARs. This presents the distribution of $\omega_{pe}$ along the loop, from which the distribution of $\omega_{pe}/\Omega_{ce}$ can be easily deduced (see Figure 1(b)). The maximum value of $\omega_{pe}/\Omega_{ce}$ is about 0.4, obtained around the loop top; the minimum is less than 0.1, obtained around the left foot. Four sections (A–D; see Figure 1(a)) are selected to infer the VDFs, and each section is given by a 10″ long tube so as to capture sufficient electrons. The average value of $\omega_{pe}/\Omega_{ce}$ is 0.3, 0.23, 0.16, and 0.1 for sections A–D, respectively. These values are much less than unity, favoring the occurrence of ECME.

Following Yousefzadeh et al. (2021), we inject energetic electrons impulsively from the loop top. To track the motion of individual electrons, the GC method (Northrop 1963;

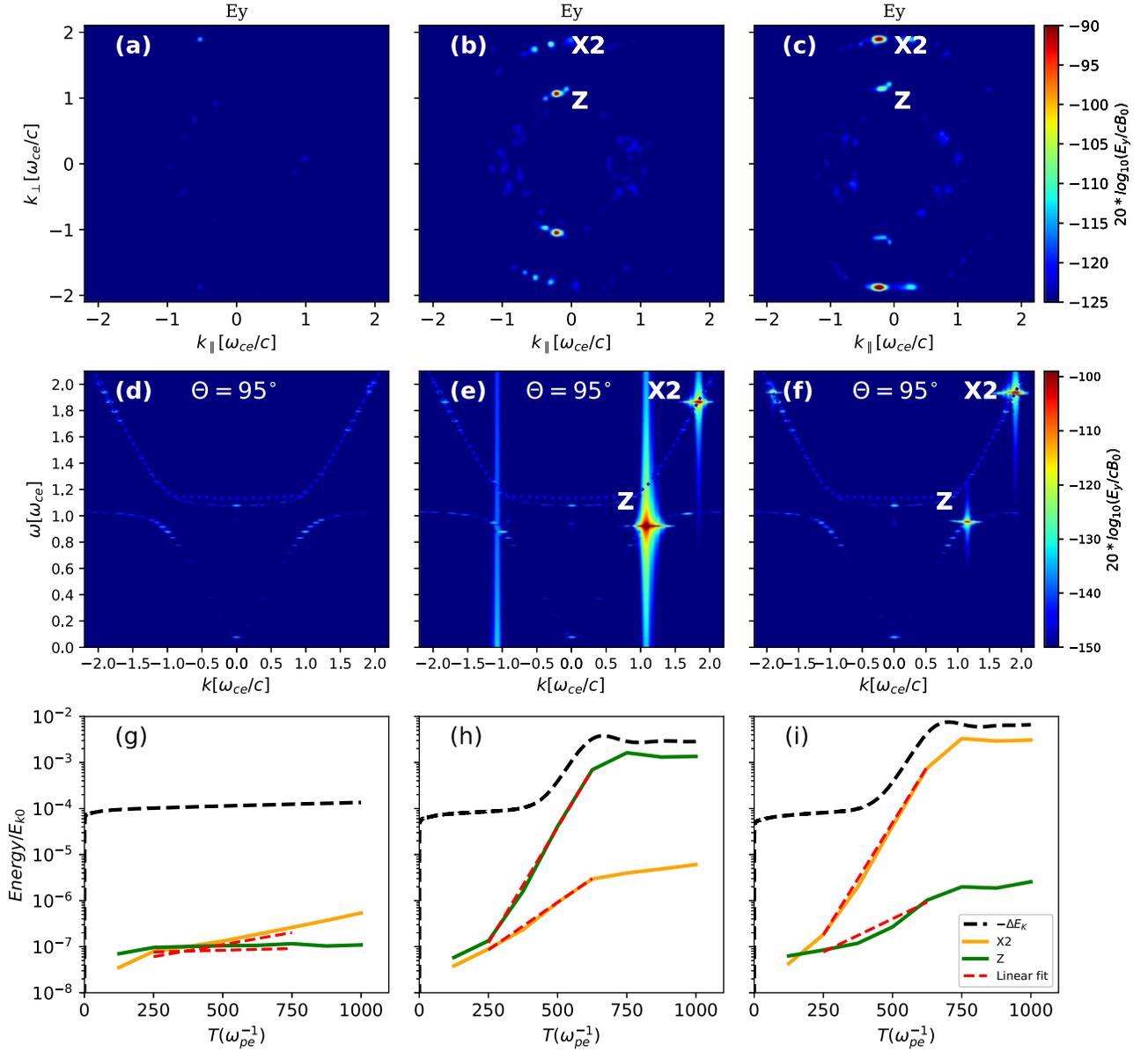

**Figure 3.** The PIC analyses for section A at 0.4 (left), 0.7 (middle), and 1 s (right) postinjection. Top panels: distribution of maximum intensity of electric field components in the $k_\parallel$–$k_\perp$ space. Middle panels: wave dispersion diagrams at propagation angles of $\theta_B = 95°$, as shown by the color map of $20 \log_{10} [Ey/(cB_0)]$. Bottom panels: temporal profiles of energies of various wave modes (X2 and Z), normalized to the total energy of energetic electrons ($E_{k0}$). Dashed lines represent exponential fitting of linear growth rates (see Table 2). The time interval of the spectral analysis is [500, 1000] $\omega_{pe}^{-1}$, and $\Delta E_k$ represents the negative variation of the electron kinetic energy.

Gordovskyy et al. 2010) was used. In total, we have tracked 4 million electrons, which follow the Maxwellian distribution with the thermal velocity being $0.24c$ (~15.6 keV). The VDFs are obtained for each section at three different times (0.4, 0.7, and 1 s postinjection) by collecting the velocity-space information of the electrons therein.

To simulate the scattering process by waves or turbulence on the electrons, following Yousefzadeh et al. (2021), the pitch angle of each electron was randomly modified once per interval ($\tau$), which represents the scattering timescale. We still refer to Chen & Petrosian (2013) to specify $\tau$, which is set to be 2 s here. This corresponds to the suggested scattering time of the ~15 keV electrons during large flares. This value is about the medium value of those specified in Yousefzadeh et al. (2021) for weakly and strongly scattering cases ($\tau = 5$ and 0.5 s, respectively).

Again, the vector particle-in-cell (VPIC; Bowers et al. 2008a, 2008b, 2009) code released by Los Alamos National Labs is used to simulate the kinetic instability or masering process within each section. This is done by initiating the simulation with the obtained VDFs of energetic electrons. The distribution of the electron–proton background plasma is taken to be Maxwellian (~2 MK), and the density ratio of energetic to background electrons ($n_e/n_0$) is taken to be 0.05.

For PIC simulations, the background magnetic field is along the $z$ direction ($B_0 \hat{e}_z$), and the wavevector ($k$) is placed in the $xOz$ plane. The domain is [512, 512] $\Delta$, where the cell size $\Delta = 2.7\, \lambda_D$, and $\lambda_D$ is the Debye length of the background electrons. For cases reaching the saturation level of wave evolution, the simulation duration is taken to be 1000 $\omega_{pe}^{-1}$; for other cases, the simulations are continued until saturation. In each cell and for each species, 1000 macroparticles are employed. The condition of charge neutrality is maintained.

Table 1
Parameters Used to Plot the Resonance Curves (see Figure 2) of Wave Excitations for Sections A–D at Different Timings

| Section | Time (s) | Mode | $\omega(\Omega_{ce})$ | $(\Omega_{ce}/c)$ | $\theta(°)$ |
|---|---|---|---|---|---|
| A | 0.4 | … | … | … | … |
|   | 0.7 | Z | 0.92 | 1.1 | 102 |
|   |     | X2 | 1.86 | 1.85 | 100 |
|   | 1.0 | Z | 0.96 | 1.1 | 95 |
|   |     | X2 | 1.92 | 1.9 | 95 |
| B | 0.4 | X2 | 1.97 | 1.99 | 105 |
|   | 0.7 | Z | 0.93 | 1.1 | 96.5 |
|   | 1.0 | Z | 0.96 | 1.1 | 95 |
|   |     | X2 | 1.91 | 1.9 | 95 |
| C | 0.4 | Z | 0.97 | 1.15 | 105 |
|   |     | X2 | 1.93 | 1.99 | 105 |
|   | 0.7 | Z | 0.91 | −0.98 | 79 |
|   | 1.0 | Z | 0.95 | 1.1 | 95 |
| D | 0.4 | Z | 0.96 | −1.1 | 75 |
|   | 0.7 | Z | 0.98 | 1.1 | 100 |
|   | 1.0 | Z | 0.96 | 1 | 90 |

Table 2
Detailed Information on the Wave Excitation for Each Section

| Section | Time (s) | Mode | Intensity ($E_{k0}$) | Growth Rate ($\omega_{pe}^{-1}$) | Significant (S) or Weak (W) | Responsible Strip Number |
|---|---|---|---|---|---|---|
| A | 0.4 | … | … | … | … | … |
|   | 0.7 | Z | 1.7e-3 | 0.02 | S | Strip 2 |
|   |     | X2 | 6.3e-6 | 0.008 | W |  |
|   | 1.0 | Z | 3.0e-6 | 0.007 | W | Strip 2 |
|   |     | X2 | 3.8e-3 | 0.021 | S |  |
| B | 0.4 | X2 | 1.8e-3 | 0.012 | S | Strip 1 |
|   | 0.7 | Z | 9.0e-4 | 0.022 | S | Strip 2 |
|   | 1.0 | Z | 2.3e-4 | 0.018 | S | Strip 2 |
|   |     | X2 | 1.9e-3 | 0.02 | S |  |
| C | 0.4 | Z | 1.2e-3 | 0.02 | S | Strip 1 |
|   |     | X2 | 6.6e-6 | 0.006 | W |  |
|   | 0.7 | Z | 6e-6 | 0.005 | W | Strip 2 |
|   | 1.0 | Z | 1.0e-3 | 0.019 | S | Strip 2 |
| D | 0.4 | Z | 2.1e-5 | 0.015 | W | Strip 1 |
|   | 0.7 | Z | 2.1e-4 | 0.009 | S | Strip 1 |
|   | 1.0 | Z | 5.8e-4 | 0.015 | S | Strip 2 |

## 3. Results of the GC Simulation: Formation of Loss-cone and Strip-like VDF Features

The VDFs for each section are obtained by plotting the phase space location of all electrons therein. See Figure 2 and the accompanying movies for VDFs at three moments (0.4, 0.7, and 1 s postinjection) and the four sections (A–D). Positive (negative) $v_\parallel$ represents electrons moving toward the right (left) foot of the loop. The Maxwellian distributions for the background electrons have been overplotted as the central dark red region.

Electrons are injected within section A (see the red squares indicated by "I"). With time going on, the electrons start to move toward both sides of the loop and appear in other sections. The VDFs contain two major components (see also Yousefzadeh et al. 2021). The first is the loss-cone component, which starts to form within a few tenths of a second and fully develops after ∼1 s for all sections due to the trap-and-loss effect of the loop structure. Note that the loss-cone angles are larger for lower sections, which have larger mirror ratios. The other is the strip-like feature. Each strip contains electrons moving bidirectionally, and the earlier one has electrons with a shorter time of mirroring. They originate from the bouncing motion of energetic electrons within the loop. Only those within certain ranges of pitch angles (satisfying the bouncing

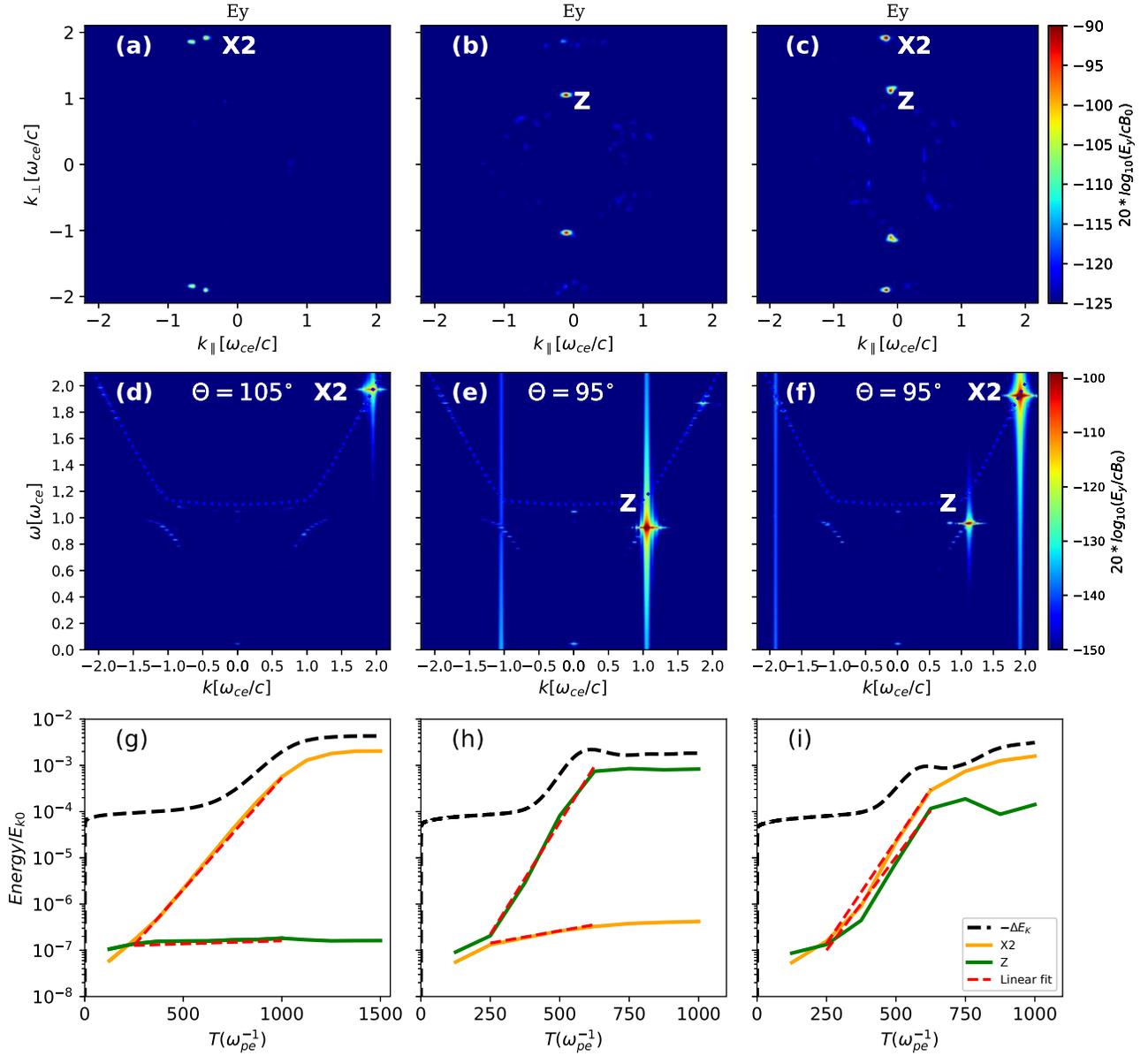

**Figure 4.** Same as Figure 3 but for section B. The wave dispersion diagrams in the middle panels are for $\theta_B = 105°$ at 0.4 s and $\theta_B = 95°$ at both 0.7 and 1 s.

condition) can be mirrored and arrive at the given location at the given time.

The number of such features and their relative intensity depend on the location of the sections and the level of scattering (Yousefzadeh et al. 2021). Four strips in section A but only two strips in section D can be identified for the first 3 s after injection. Later strips appear less sharp with fewer electrons and a smaller gradient of the VDF due to the continuous scattering and loss of electrons. For the lower part of the loop (sections C and D), the second strip appears later and is less intense. According to Yousefzadeh et al. (2021), if considering a stronger scattering effect, the features get more obscure and even disappear completely.

All strips move (with time) toward the bottom left regime of the ($v_\perp$–$v_\parallel$) space as a whole. This means that electrons arriving at each section earlier have a generally larger $v_\perp$ and $v_\parallel$. The strips eventually get mixed with the loss-cone feature. Note that the appearance of the strip-like feature of the VDF was first suggested by White et al. (1983) based on an analytical solution of the simplified Boltzmann equation along a magnetic arch structure (see their Figure 2).

## 4. Results of the PIC Simulation: Excitation of the Harmonic Cyclotron Maser Emission

The PIC simulations start from the VDFs of energetic electrons obtained at 0.4, 0.7, or 1 s (postinjection time) combined with the background VDF as presented in Figure 2. With the same method, Yousefzadeh et al. (2021) investigated the masering effect of the loss-cone and strip features of the VDF deduced from the top section (A) at 1 s. They found that the strip feature can excite the quasi-perpendicular propagating harmonic X (X2) and Z mode, while the loss-cone feature can excite the fundamental X (X1) mode. Here, for comparison, we set the scattering timescale ($\tau$) to be 2 s; as mentioned, it is about the median value of those for weakly and strongly scattering cases (referred to as case W with $\tau = 5$ s and case S with $\tau = 5$ s in Yousefzadeh et al. 2021). Similar to their case

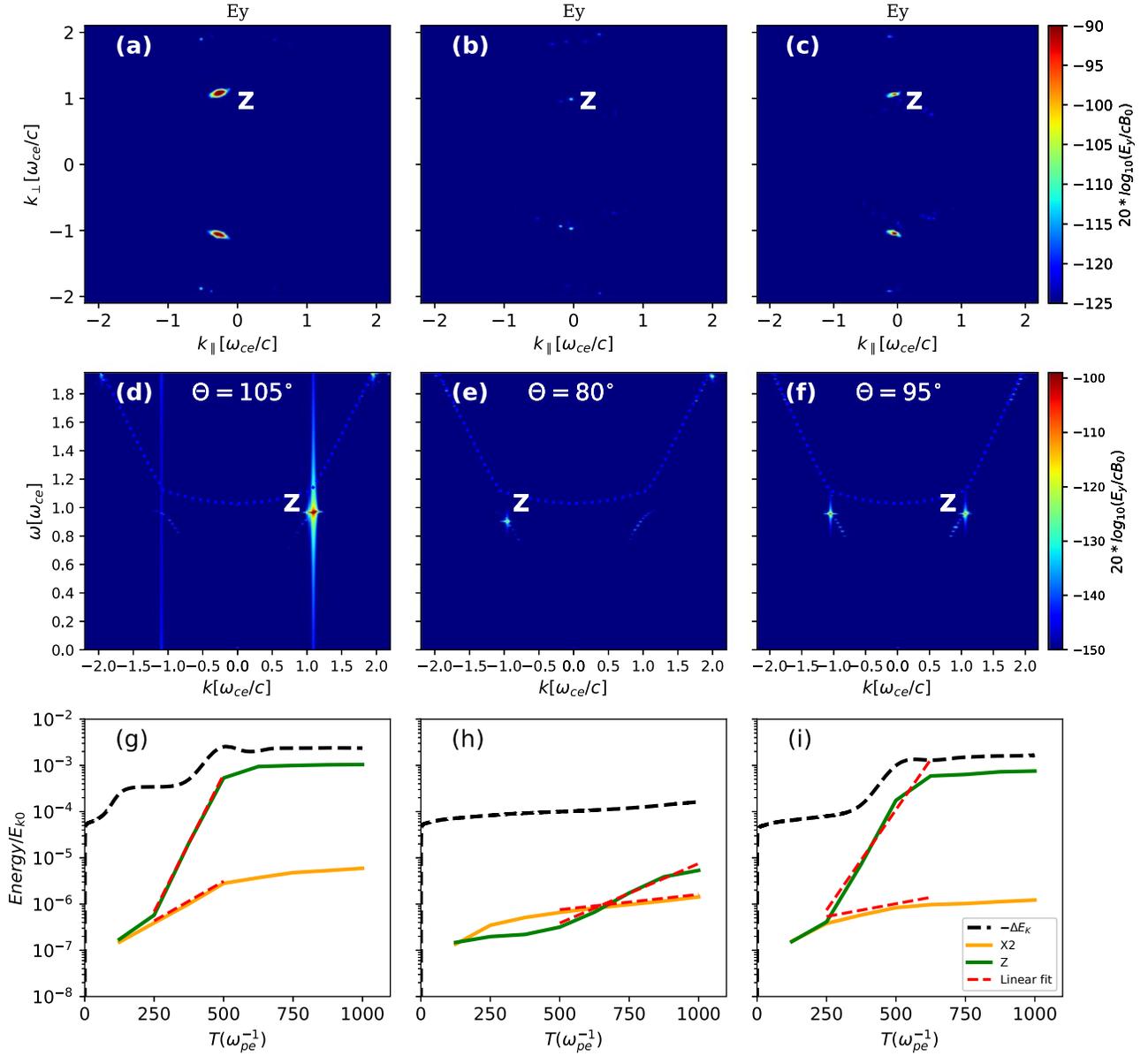

**Figure 5.** Same as Figure 3 but for section C. The wave dispersion diagrams in the middle panels are for $\theta_B = 105°$, $80°$, and $95°$ at 0.4, 0.7, and 1 s, respectively.

W, we observe no significant excitation of X1 for all cases presented here due to the relatively weak scattering effect and thus the underdeveloped loss-cone distribution at the time of interest ($\leqslant 1$ s). Therefore, we do not discuss the X1 mode hereafter.

The excitation levels of the two modes, X2 and Z, can be very different, depending on timing and location along the loop. The levels are classified into two groups: "significant" for intensity reaching above $1. \times 10^{-4} E_{k0}$ and "weak" for those below this value. Starting from section A, i.e., the top part of the loop, for the VDF obtained at 0.4 s (referred to as the 0.4 s VDF), we observe no significant emission of any mode. For the 0.7 s VDF, significant excitation of Z and weak emission of X2 can be observed, with the Z-mode intensity reaching $\sim 1.7 \times 10^{-3} E_{k0}$, and both are quasi-perpendicular (see Figure 3). For the 1 s VDF, we observe excitation of weak Z and strong X2 with the latter reaching $3.8 \times 10^{-3} E_{k0}$. To figure out the energy source of the excitation, we examine the resonance curves given by the parameters ($\omega$, $k$, $\theta$) of the excited mode (see Table 1). For both modes and for VDFs obtained at both moments (0.7 and 1 s), the second strip (or strip 2, indicated by the blue arrow) is the responsible VDF feature, while the first strip (also called the crescent feature; see, e.g., Vorgul et al. 2011 and Wu et al. 2012) plays no role, since it passes through the central part of the background distribution. These pieces of information, including the exciting agency (strips 1 or 2; see arrows in Figure 2(e)), are listed in Table 2.

For section B, the 0.4 s VDF results in significant quasi-perpendicular excitation of X2 (see Figure 4). The X2 excitation is due to the first strip, which deviates away from the center of the background distribution as seen from Figure 2(d), but without observable excitation of the Z mode. For the 0.7 s VDF, the X2 mode manifests weak excitation, while the Z mode is excited significantly, with the intensity reaching $\sim 9.0 \times 10^{-4} E_{k0}$. Both are due to strip 2. For the 1 s VDF, both X2 and Z are excited significantly, also due to strip 2.

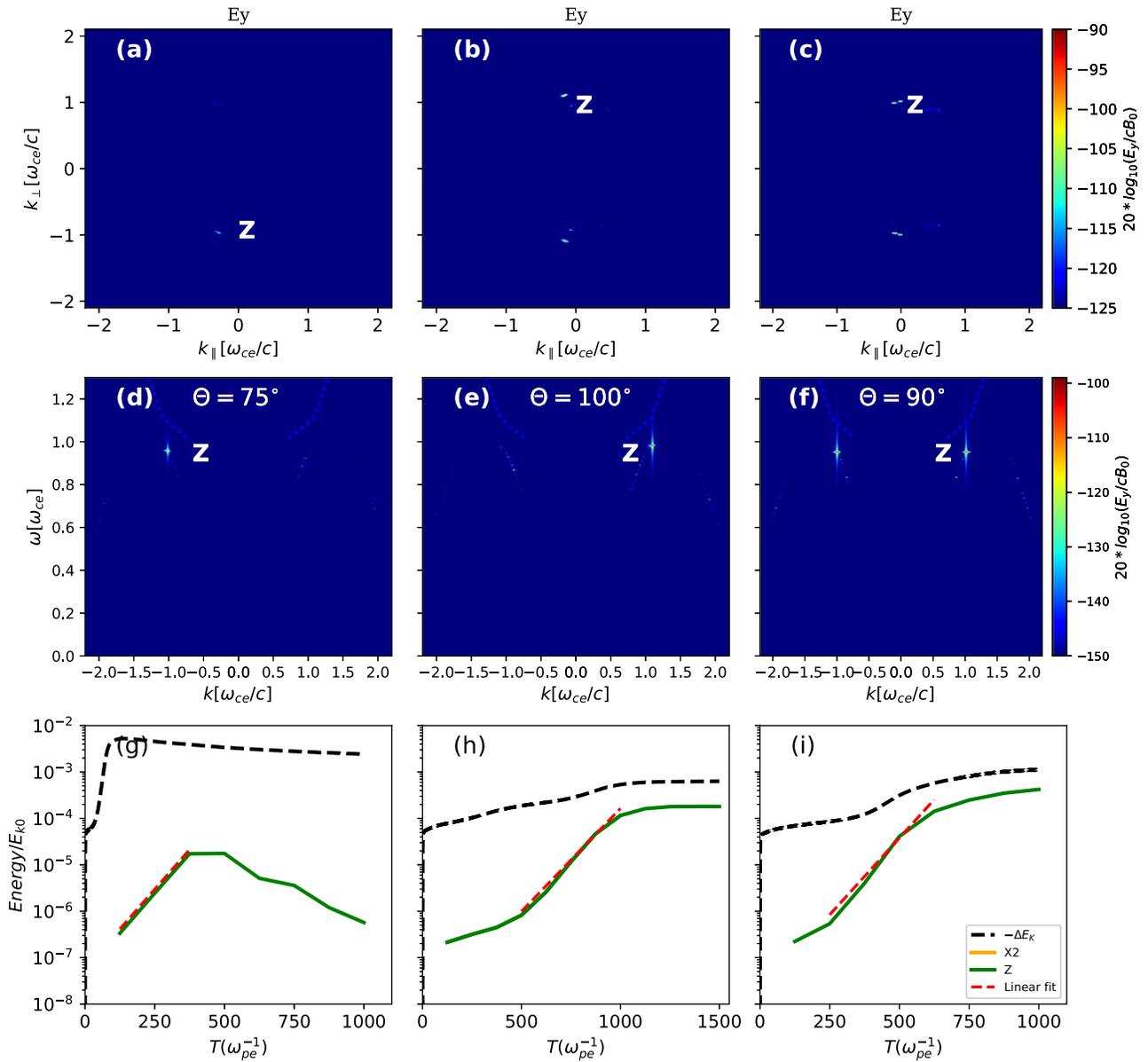

**Figure 6.** Same as Figure 3 but for section D. The wave dispersion diagrams in the middle panels are for $\theta_B = 75°$, $100°$, and $90°$ at 0.4, 0.7, and 1 s, respectively.

The VDFs associated with sections C and D always lead to weak excitation of X2 (see Figures 5 and 6), i.e., its intensity is less than $1 \times 10^{-4} E_{k0}$, while the Z mode can be excited significantly by strip 1 for the 0.4 s VDF and strip 2 for the 1 s VDF within section C and by strip 2 for the 1 s VDF within section D. See Table 2 for a summary of the relevant information of wave excitation.

According to the movie accompanying Figure 2, strip 2 continues to move toward and get mixed with the background distribution after 1 s of injection. Thus, the associated VDF gradient and the rate of wave excitation shall decrease correspondingly.

## 5. Summary and Discussion

Following Yousefzadeh et al. (2021), we continue to investigate the ECME process along the selected loop structure with the three-step numerical scheme designed to consider the multiscale process underlying solar radio bursts. The main purpose is to further explore the masering process along the loop at different locations and timings in response to the impulsive injection of energetic electrons from the loop top. The study focuses on the mechanism of harmonic X2 emission so as to resolve the escaping difficulty of the ECME theory when it is applied to solar radio bursts, such as spikes. The loop structure is represented by four sections, within which the VDFs are obtained by the GC of particle motion. The VDFs contain two main components, the loss-cone and strip-like components, consistent with our earlier study. The kinetic instability giving rise to the emitting process is investigated with fully kinetic electromagnetic PIC simulations fed by the obtained VDFs at different loop sections and timings.

Significant emission of the Z mode and no significant emission of X1 are observed for all loop sections. Strong excitations of X2 are only from the upper part of the loop (sections A and B), given by both strips of VDF. They first come from the upper middle part of the loop (section B) due to the first strip of the 0.4 s VDF (i.e., the VDF obtained at 0.4 s postinjection), and then along the whole upper part of the loop (sections A and B) due to the second strip of the corresponding 1 s VDF; the strongest X2 emission is from the top section (section A) due to the second strip of the 1 s VDF reaching an intensity of $3.8 \times 10^{-3} E_{k0}$.

The present study, together with those of Yousefzadeh et al. (2021) and Ning et al. (2021a, 2021b), provide a novel possibility of efficient excitation of X2 via ECME. This also provides a practical means to circumvent the escaping difficulty met by earlier studies of the ECME theory with the X1 mode being the primary radiation (see, e.g., Wu et al. 2002; Yoon et al. 2002). This is critical to understanding the origin of solar radio bursts, such as spikes, and other astrophysical radio bursts that might be explained with ECME.

The X2 mode is mainly excited by the transient strip-like features of the VDF of energetic electrons that are formed during their transport through the magnetic mirror structure, rather than the more general loss-cone feature. Note that the two features can coexist in the source of radio emission; thus, it should not be taken for granted that the loss-cone maser is the radiation mechanism if the loss-cone distribution is detected. The strip-like features may be a common phenomenon during solar flares, which are very similar to the shell-like or so-called horseshoe distribution that is responsible for the auroral kilometric radiation. Ning et al. (2021b) investigated the maser emission driven by energetic electrons with the horseshoe distribution using the same PIC simulation code; they also found efficient excitation of X2, and the rate of energy conversion from energetic electrons into X2 varies from 0.06% to 0.17%, depending on the relative abundance of energetic electrons (varying from 5% to 10%). This is in line with the present result.

According to our simulations, the X2 emission mainly comes from the upper part of the selected coronal loop and grows after a few tenths of a second after the injection of heated electrons; it may get intensified intermittently, reaching high intensity around 0.4 or 1 s postinjection. If considering the complexity of the magnetic structure (e.g., multiple threads of loops) and the intermittent nature of electron acceleration and injection due to reconnection, the radio intermittency will become much more apparent. This is in line with the observed intermittency of solar radio bursts, such as spikes.

Solar spikes are characterized by a very short lifetime, very narrow bandwidth, strong polarization, and very high brightness temperature ($\geqslant 10^{11}$–$10^{15}$ K) in metric-decimetric wavelengths. They tend to occur in great numbers (up to thousands; e.g., Benz 1985, 1986; Fleishman & Melnikov 1998; Benz et al. 2002; Rozhansky et al. 2008; Tan 2013). Solar spikes are suggested to be the elementary bursts of a solar flare.

To apply PIC simulation results to solar spikes, however, is not straightforward since huge differences of temporal and spatial scales exist between modelings and observations. Observational resolutions are usually $\geqslant 1$ ms and $> 10^4$ km, while modeling scales are less than 1 $\mu$s and tens of kilometers. The differences are in several orders of magnitude. The situation becomes even worse if the unknown propagation effect needs to be considered. Nevertheless, in the following, we still try to predict some observational characteristics based on the simulations.

First, the modelings clearly indicate a high polarization (nearly 100%) of the obtained harmonic emission; the emission frequency is about 1–3 GHz, and the relative bandwidth is ∼0.06. The emission propagates almost perpendicularly to the background field within a narrow angular range of less than 10°. This means that only less than 1/36 of the bursts could propagate toward the observer (neglecting scattering and propagation effects), accounting for the observation that only a few percent of hard X-ray bursts are associated with spikes. To estimate $T_B$, we refer to Winglee & Dulk (1986) for the appropriate equations (see Ning et al. 2021a for a similar estimate). The $T_B$ is estimated to be ∼$10^{15}$–$10^{16}$ K (for strong cases). These predictions are in line with major observations of solar spikes. The highly dynamic and intermittent nature of energetic electrons in the aftermath of intermittent and sporadic flaring reconnections may account for the group occurrence of spikes in great numbers.

In our configuration of simulations, the selected loop represents magnetic structures that get accessed by energetic electrons from the top part during solar flares. The loop could be a newly reconnected flare loop. These energetic electrons may have VDFs that are very different from Maxwellians; for instance, they may take a beam-like distribution or drifting Maxwellian distribution or others, depending on the acceleration mechanism of reconnection. The effect of different VDFs, locations of injection, and coronal conditions shall be investigated in future.

The study is supported by NNSFC grants (11790303, 11790300, 11973031, and 11873036). The authors acknowledge the open-source Vector Particle In Cell (VPIC) code provided by Los Alamos National Labs (LANL) and the Super Cloud Computing Center (BSCC; http://www.blsc.cn/) for providing computational resources.

### ORCID iDs

Mehdi Yousefzadeh  https://orcid.org/0000-0003-2682-9784
Yao Chen  https://orcid.org/0000-0001-6449-8838
Hao Ning  https://orcid.org/0000-0001-8132-5357